\documentclass[prb,preprint,aps,superscriptaddress]{revtex4}

\usepackage{graphicx}
\usepackage{color,soul}
\usepackage{url}

\begin{document}

\title{Topological Node-Line Semimetal in Three Dimensional Graphene Networks}

\author{Hongming Weng}

\email{hmweng@iphy.ac.cn}

\affiliation{Beijing National Laboratory for Condensed Matter Physics,
  and Institute of Physics, Chinese Academy of Sciences, Beijing
  100190, China}

\affiliation{Collaborative Innovation Center of Quantum Matter,
  Beijing, China}

\author{Yunye Liang}

\affiliation{New Industry Creation Hatchery Center, Tohoku University,
  Sendai, 980-8579, Japan}

\author{Qiunan Xu}

\affiliation{Beijing National Laboratory for Condensed Matter Physics,
  and Institute of Physics, Chinese Academy of Sciences, Beijing
  100190, China}

\author{Rui Yu}

\affiliation{International Center for Materials Nanoarchitectonics (WPI-MANA),
National Institute for Materials Science, Tsukuba 305-0044, Japan}

\author{Zhong Fang}


\affiliation{Beijing National Laboratory for Condensed Matter Physics,
  and Institute of Physics, Chinese Academy of Sciences, Beijing
  100190, China}

\affiliation{Collaborative Innovation Center of Quantum Matter,
  Beijing, China}

\author{Xi Dai}


\affiliation{Beijing National Laboratory for Condensed Matter Physics,
  and Institute of Physics, Chinese Academy of Sciences, Beijing
  100190, China}

\affiliation{Collaborative Innovation Center of Quantum Matter, Beijing, China}

\author{Yoshiyuki Kawazoe}

\affiliation{New Industry Creation Hatchery Center, Tohoku University,
  Sendai, 980-8579, Japan}

\affiliation{Thermophysics Institute, Siberian Branch, Russian Academy
  of Sciences, Russia}

\date{\today}

\begin{abstract}
  Graphene, a two dimensional (2D) carbon sheet, acquires many of its
  amazing properties from the Dirac point nature of its electronic
  structures with negligible spin-orbit coupling. Extending to 3D
  space, graphene networks with negative curvature, called
  Mackay-Terrones crystals (MTC), have been proposed and
  experimentally explored, yet their topological properties remain to
  be discovered. Based on the first-principle calculations, we
  report an all-carbon MTC with topologically non-trivial electronic
  states by exhibiting node-lines in bulk. When the node-lines 
  are projected on to surfaces to form circles, ``drumhead"-like 
  flat surface bands nestled inside of the circles are formed. 
  The bulk node-line can evolve into 3D Dirac 
  point in the absence of inversion symmetry, which has shown its plausible 
  existence in recent experiments.   
\end{abstract}

\maketitle

\section{introduction} \label{introduction}
Carbon is one of the most fascinating elements in nature, which can
form many different crystal structures with diverse electronic
properties, such as C$_{60}$,~\cite{C60} nanotube,~\cite{nanotube} 
graphene,~\cite{Graphene-scotch} graphite, and
diamond. Among them, graphene is one of the most amazing materials,
which supports Dirac point in its low energy electronic structure,
described as $H= v \vec{k} \cdot \vec{\sigma}$, where $v$ is velocity,
$\vec{k}=(k_x, k_y)$ is momentum and $\vec{\sigma}$ is Pauli
matrix. This novel electronic state leads to many interesting
phenomena, such as the unconventional quantum Hall effect, large
magnetoresistance, and unusual optical properties, which make graphene
potentially useful.~\cite{Graphene} The presence of 2D Dirac cone is fragile, and two
conditions are required to protect it: (1) the absence of spin-orbit
coupling (SOC), and (2) the presence of inversion symmetry. The first
condition is naturally satisfied in graphene, because its SOC strength
is negligible ($\sim 10^{-3}$ meV).~\cite{Yao} If SOC in graphene is taken into account, it
will open up a gap at Fermi level and lead to quantum spin Hall
insulator (i.e., 2D topological insulator).~\cite{Kane} The second requirement is,
however, very strong, and it is satisfied only in the presence of A-B
sublattice symmetry, which can be easily broken, leading to a normal
insulating state, similar to that in BN nanosheet.

As proposed by A. L. Mackay and H. Terrones,~\cite{mackay} graphene can 
be extended to three-dimensional (3D) space to form 3D networks by 
placing graphitic tiles consisting of 4- to 8-membered rings onto 
the Schwarz minimal surface. Hereafter, we call such 3D all carbon allotrope 
Mackay-Terrones crystal (MTC). Schwarz minimal surface is a 3D periodic 
minimal surface with its mean curvature $H=(k_1+k_2)/2$ being zero 
and Gaussian curvature ($K=k_1k_2$) being negative everywhere on it. 
Here $k_1$ and $k_2$ are the principal curvatures. There are various 
Schwarz minimal surfaces, such as primitive (P), diamond (D) and gyroid (G) 
surface. One type of MTC based on P-surface is shown in Fig. 1. 
Different from C$_{60}$-like fullerene, which has positive Gaussian 
curvature, MTC has negative Gaussian curvature and is periodically 
connected. Such 3D network of $sp^2$-bonded 
carbon has unique properties, such as high surface-to-volume ratio and 
remarkable porosity, which stimulate extensive studies.~\cite{view, bio} 
Theoretically, MTC has been proved to be dynamically 
stable and require less formation energy than C$_{60}$.~\cite{dft_mackay, vanderbilt} 
Experimentally, saddle-like nano-carbon sheet, the main component of MTC, 
has  been successfully synthesized.~\cite{saddle} Similar negatively 
curved $sp^2$ networks has been observed in spongy
carbon~\cite{spongy} and negative replica of zeolite.~\cite{zeolite} Recently, high-quality 3D 
nanoporous graphene fabricated by using nanoporous Ni as template 
shows very similar MTC structure,~\cite{chenlab1, chenlab2} 
making synthesizing of it very promising. On the other hand, the topological 
properties of the band structure for these all-carbon MTC remain 
unexplored and will be the main subject of this paper. We will show that 
such all-carbon MTC can host non-trivial electronic states, including 
topological node-lines and 3D Dirac points, which are distinct from its 2D counter 
material graphene.

\section{Results} \label{Results}
We concentrate on the MTC formed with Schwarz minimal P-surface. 
As shown in Fig. 1, a stable structure with simple cubic lattice in
$Pm\bar{3}m$ space group, and 176 atoms per unit cell has been obtained
by Tagami {\it et al.} in Ref.~\onlinecite{ours} and labeled as 6-1-1-p. We 
have employed the software package OpenMX~\cite{openmx} for the 
first-principles calculation. It is based on norm-conserving pseudopotential
 and pseudo-atomic localized basis functions, which is highly 
 efficient for the MTC with more than a hundred atoms. The choice of 
 pseudopotentials, pseudo atomic orbital basis sets C6.0-s2p2d1 and 
the sampling of Brillouin zone with $10\times10\times10$-grid have been 
carefully checked. After full structural relaxation, we get the lattice
constant $a$=14.48 \AA, and the diameters of the pipes or pores are around
9.876 \AA$ $ and 5.629 \AA, respectively, which are in good agreement 
with their results.~\cite{ours} The electronic band structure of this crystal, 
calculated based on the local density approximation (LDA), is shown in
Fig. 1(d). We find that this crystal is a semimetal with 
band crossings around the Fermi level, similar to the massless Dirac cone in
graphene, but they are in fact very different -- the key issue of this paper.

\subsection{Band structure}
Detailed analysis of the band structure reveals that: (1) The occupied
and unoccupied low energy bands are triply-degenerate at $\Gamma$, and
have T$_{1g}$ and T$_{2u}$ symmetry, respectively. The formers are
even while the laters are odd under spacial inversion symmetry. Moving from
$\Gamma$ to R point, their degeneracy recover again.
However, their energy ordering exchanges, leading to the so called
band inversion, which is one of the key ingredients for the
topological insulators.~\cite{TIreview, TIreview-2} Due to the band
inversion, the band crossings happen along both $X-R$ and $R-M$
paths as seen from Fig. 1(d). (2) Including SOC in the
calculation, a gap will open up around the band crossings, leading to
a 3D strong topological insulator with $Z_2$ index of (1;111)~\cite{Z2} 
by treating the lower half of the anti-crossing bands as occupied. 
However, similar to graphene, the SOC splitting is small (around 0.13 meV or 1.5 K), 
and can be neglected in cases with temperature higher than 1.5 K.

\begin{figure}
\includegraphics[scale=0.6]{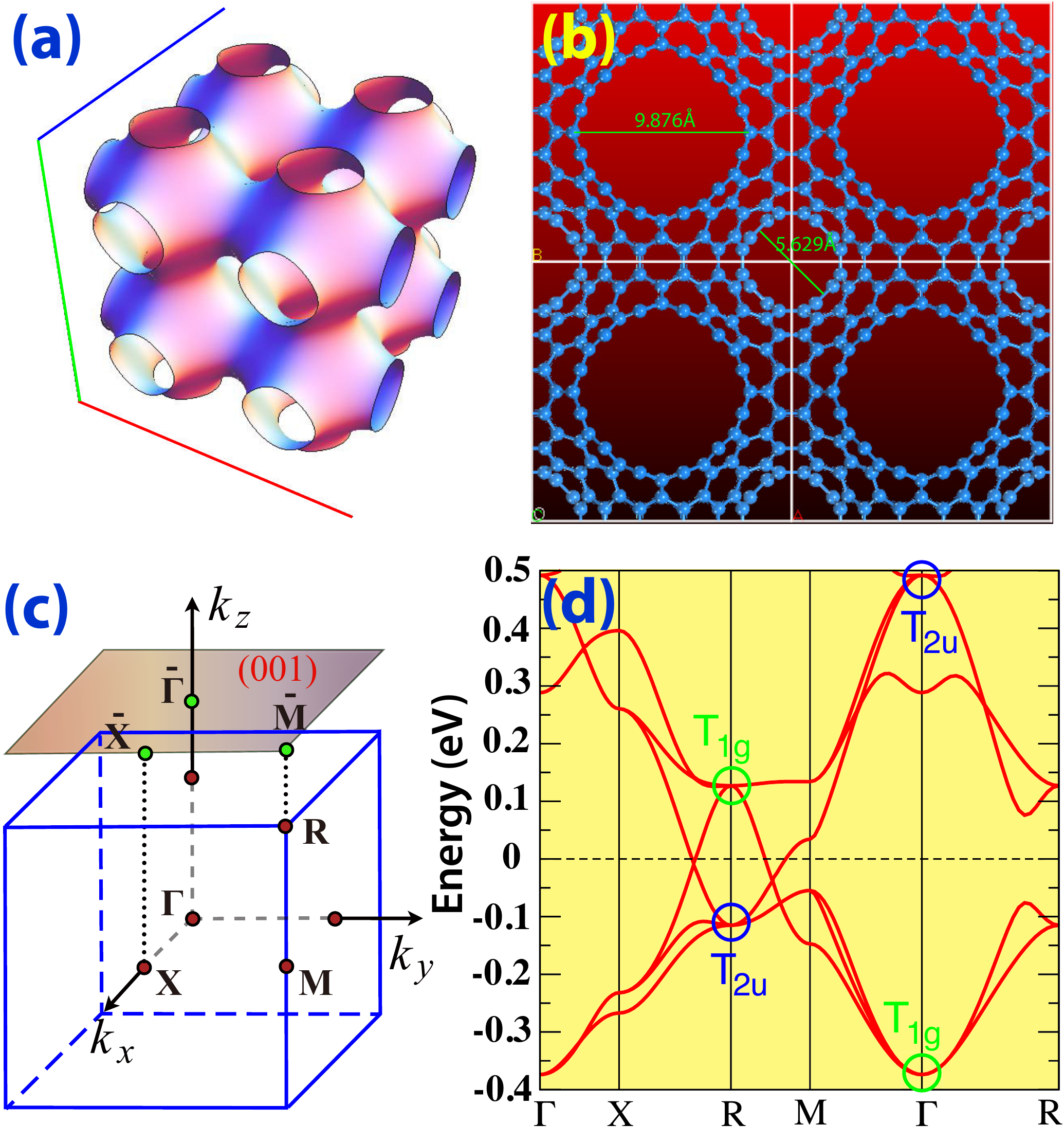}
\caption{(Color online) (a) The Schwarz minimal P-surface in 2$\times$2$\times$2 supercell. (b) The top view of 
6-1-1-p MTC in 2$\times$2 supercell. (c) Bulk and (001)-surface Brillouin zone, as well as 
the high symmetrical k-points. (d) Band structure from the first-principles calculation. The two triply degenerated 
eigenstates at $\Gamma$ and R with T$_{1g}$ and T$_{2u}$ symmetrical representation are marked. The band 
inversion between them can be easily seen.}
\label{MTCstructure}
\end{figure}

\begin{figure}
\includegraphics[scale=0.6]{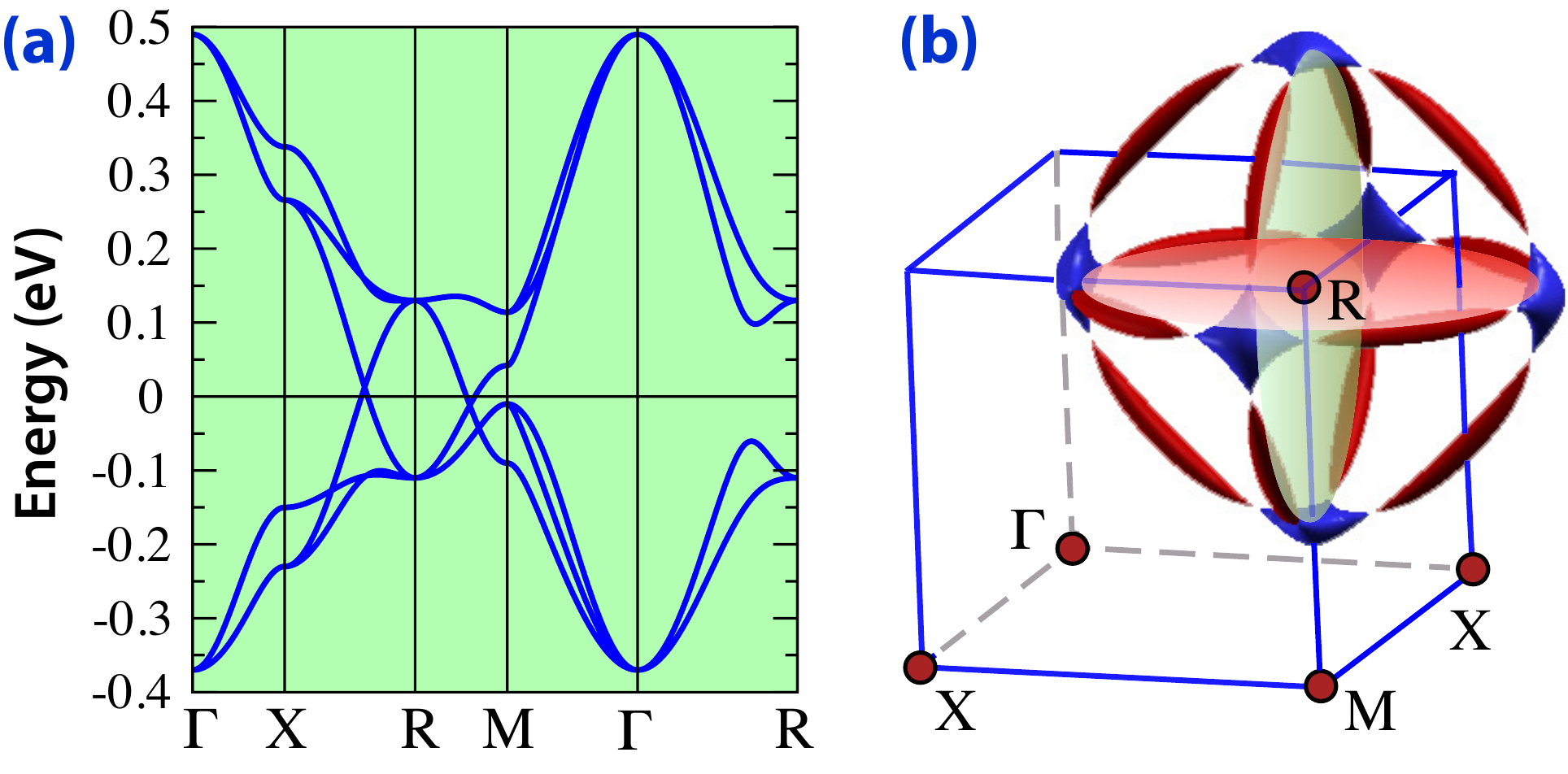}
\caption{(Color online) (a) Band structure from effective tight-binding model calculation, which reproduces all the features 
of Fig. 1(d). (b) Fermi surface consists of three lotus root like rings. These rings are centering R point 
and parallel to the $k_x$=$\frac{\pi}{a}$, $k_y$=$\frac{\pi}{a}$ and $k_z$=$\frac{\pi}{a}$ plane, respectively. They are formed by
the electron pockets (blue) and hole pockets (red) connected by nodal points at Fermi energy. 
}
\label{tbband}
\end{figure}

The low energy bands near the Fermi level are formed by 
the overlapping of the molecular orbitals with T$_{1g}$ and T$_{2u}$ symmetry. 
Since each isolated carbon cluster in the MTC has approximately spherical symmetry,
these molecular orbitals can be viewed as the ``atomic orbitals" with  $g$ and $f$-wave
symmetry, which are further splitted under the cubic crystal field.
Therefore, the T$_{1g}$ sector consists of $g_{xy(x^2-y^2)}$, 
$g_{yz(y^2-z^2)}$ and $g_{zx(z^2-x^2)}$ orbitals, while the T$_{2u}$ sector
contains $f_{x(y^2-z^2)}$, $f_{y(z^2-x^2)}$, and $f_{z(x^2-y^2)}$ orbitals.
Thus, these six hypothetical atomic orbitals are used as basis set to reproduce 
the low energy physics of this system. A Slater-Koster tight-binding (TB) Hamiltonian 
has been established and the on-site energy levels, as well as hopping
parameters, can be obtained by fitting the band structure from the first-principles calculations. 
(see Appendix for details) The triply degenerated T$_{1g}$ and T$_{2u}$ bands 
at $\Gamma$ have eigen energies of $E_g+4V_{ggp}+2V_{ggd}$
and $E_f-4V_{ffd}$, respectively. Those at R are $E_g-4V_{ggp}-2V_{ggd}$
and $E_f+4V_{ffd}$ due to the nearest-neighbor hopping. Here, 
$E_g$ and $E_f$ are on-site energies for $g$ and $f$ orbitals. $V_{ggp}$ and $V_{ggd}$ 
are the hopping parameters among $g$ orbitals. $V_{ffp}$ and $V_{ffd}$ are those for 
$f$-orbitals. From these analysis, we learn that the band inversion or the
switching of $g$ (T$_{1g}$) and $f$ (T$_{2u}$) orbitals
between $\Gamma$ and R points is due to the strong
energy dispersion (or the large hopping parameters). As shown in Fig. 2, this TB model 
can reproduce all the features of bands, including band topology, with the fitted Slater-Koster 
parameters (in unit eV) $E_g$=-0.12, $E_f$=0.19, $V_{ffp}$=0.019, $V_{ffd}$=-0.075,
$V_{fgp}$=0.05, $V_{fgd}$=0.0, $V_{ggp}$=-0.035, $V_{ggd}$=-0.055. The mean square error
is minimized to 0.0016 eV$^2$ with sampling k-points along the high-symmetrical path shown
in Fig. 1(d). Artificially reducing the hopping parameters (such as expanding the lattice
parameter) by 50\% will eliminate the band inversion, with T$_{1g}$ 
states lower than the T$_{2u}$ states at R point. This calculation also 
suggests that the strength of band inversion in the system is strong.

\subsection{Topological Node-Lines and 3D Dirac Points}

Interestingly, the band crossings in MTC lead to node-lines rather than node
points. In other words, the band crossings exist along certain closed
loops in the 3D momentum space, and generate three circular-like
node-lines around the $R$ point, as shown in Fig. 2. 
These node-lines are protected by two factors, one is the coexistence of
time reversal (T) and spacial inversion (P) symmetry and the other factor 
is that the SOC is negligible.

With the coexistence of P and T symmetries, there exists a certain gauge choice under which
the spineless Hamiltonian is completely real valued. (See Appendix for details) 
Now we will show that for this system, if there is an energy level-crossing of two bands at a 
momentum $\mathbf{k}_0$, a stable node-line will unavoidably appear.
Around the crossing point, the two-level 2$\times$2 Hamiltonian can be 
written in the following general form:
 \begin{equation}
 \mathcal{H}=d_0(\vec{k})+d_{x}(\vec{k}) \cdot \sigma_{x}+d_{y}(\vec{k}) \cdot \sigma_{y}+d_{z}(\vec{k}) \cdot \sigma_{z},
 \end{equation}
 where the Pauli matrices $\sigma_{i}$($i$=$x, y, z$) denote the
 two-band space. Without loss of generality,
 $d_{i}(\vec{k})$($i$=$0,x,y,z$) are all real function of
 $\vec{k}$.  The eigen energy of $\mathcal{H}$ being
 \begin{equation}
 E(\vec{k})=\pm\sqrt{d_{x}^2(\vec{k})+d_{y}^2(\vec{k})+d_{z}^2(\vec{k})}+d_0(\vec{k}),
 \end{equation}
 and the energy degeneracy can be obtained when the three conditions
 $d_{i}(\vec{k})$=0 ($i$=$x,y,z$) are satisfied with three
 parameters $\vec{k}(k_x,k_y,k_z)$ in the 3D momentum space.
As mentioned above, the Hamiltonian can be chosen to be real valued 
 leading to  $d_y=0$. The rest $d_{0}(\vec{k})$, $d_{x}(\vec{k})$ and $d_{z}(\vec{k})$ can be
 expanded around $k_0$ and
 the location of the crossing points can be determined by $d_{x}(\vec{k}_0)\approx \delta_x+{{\vec v_x}}(\vec{k}-k_0)=0$ and
  $d_{z}(\vec{k}_0)\approx \delta_y+{{\vec v_z}}(\vec{k}-k_0)=0$, where ${\vec v_i}=\vec\nabla_{\vec{k}} d_i(\vec{k})$ and
  $\delta_i$ denote the small perturbative terms with both T and P symmetries.
  In the generic case, the above two equations give a line in the vicinity of $k_0$ with its direction determined by ${\vec v_x}\times{\vec v_z}$.
  Therefore, the generic solution of the band crossing point in 3D k-space is a closed loop. Any external perturbations that 
  keep T, P and translational symmetry can only shift or distort but not eliminate the nodal loops.

 
The topologically stable node-line in MTC
 is only protected by P and T and no other symmetry is required. The additional mirror symmetry 
 in the present system only forces the node-lines to stay in $k_{z}$ (or $k_x, k_y$)=$\frac{\pi}{a}$ plane. 
 The cubic symmetry leads to three in-plane node-lines, as what 
 have been found from our calculations in Fig. 2. 
 The node-lines are not necessarily to be flat in energy,
 and they can have energy dispersion in the k space determined
 by $d_0({\bf k})$ term (which breaks the particle-hole
 symmetry). Different from other proposals for the topological
 node-lines,~\cite{burkov} the appearance of node-lines in MTC
 is very stable and does not require fine tuning of any parameters. 
 This mechanism to generate topological node-lines in three dimensional materials
 only requires T, P symmetry and weak enough SOC, which can be easily applied to
 a large class of materials consisting of mainly the light elements.


 It is now clear that this 3D MTC is different with graphene in the
 sense that it is a semimetal with node-lines in the 3D momentum
 space with the presence of both T and P symmetries. The thing 
 becomes even more interesting if P symmetry is further broken. In such a case,
 from above discussions, we will in general expect three conditions
 $d_i({\bf k})=0$ with three parameters for the band crossing points, 
 leading to isolated points in the 3D k-space. This is nothing but the 3D Dirac metals
 discussed recently.~\cite{Na3Bi, Cd3As2, Na3Biexp, Cd3As2exp,Cd3As2expHasan, Cd3As2expCava} 
 On the other hand, comparing with other proposals for Dirac semimetals, 
 the 3D Dirac point here is topologically stable and does not require the protection from any
 crystalline symmetry. Similar with the situation in graphene, finite SOC will open a gap at the Dirac point
 and makes the system a topological insulator.
 In fact, although our calculated structure has inversion
 symmetry, most of known real samples of MTC~\cite{chenlab1,chenlab2} 
 have strong defects and orientation disorder, which should break inversion symmetry. 
 The plausible existence of these stable 3D Dirac points has been indicated 
 by the density of states~\cite{chenlab1} and heat capacity measurements.~\cite{privatecommuniction} 
 If T symmetry is further broken in the system, we will
 expect Weyl semimetal states, which has been extensively studied but
 not realized yet experimentally.~\cite{wan, HgCrSe,multilayerTRI, multilayerTRB}
 
\begin{figure}
\includegraphics[scale=0.6]{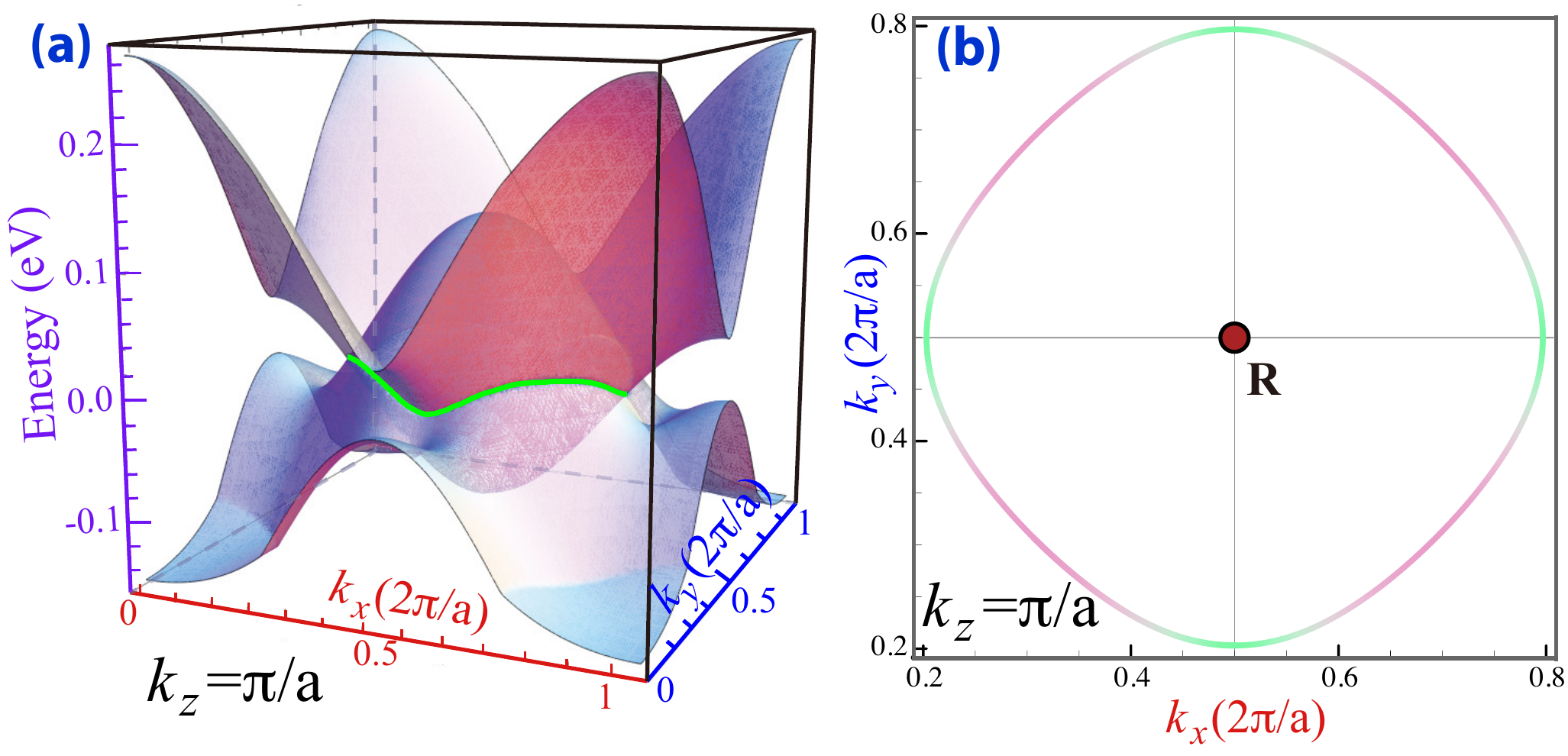}
\caption{(Color online) (a) Band crossings of the two bands near Fermi level form node-line (in Green) in 
$k_z$=$\frac{\pi}{a}$ plane. (b) The crossing happens at different eigen energy as indicated by different color, 
the greener the lower in energy. 
}
\label{nodeline}
\end{figure}

\begin{figure}
\includegraphics[scale=0.5]{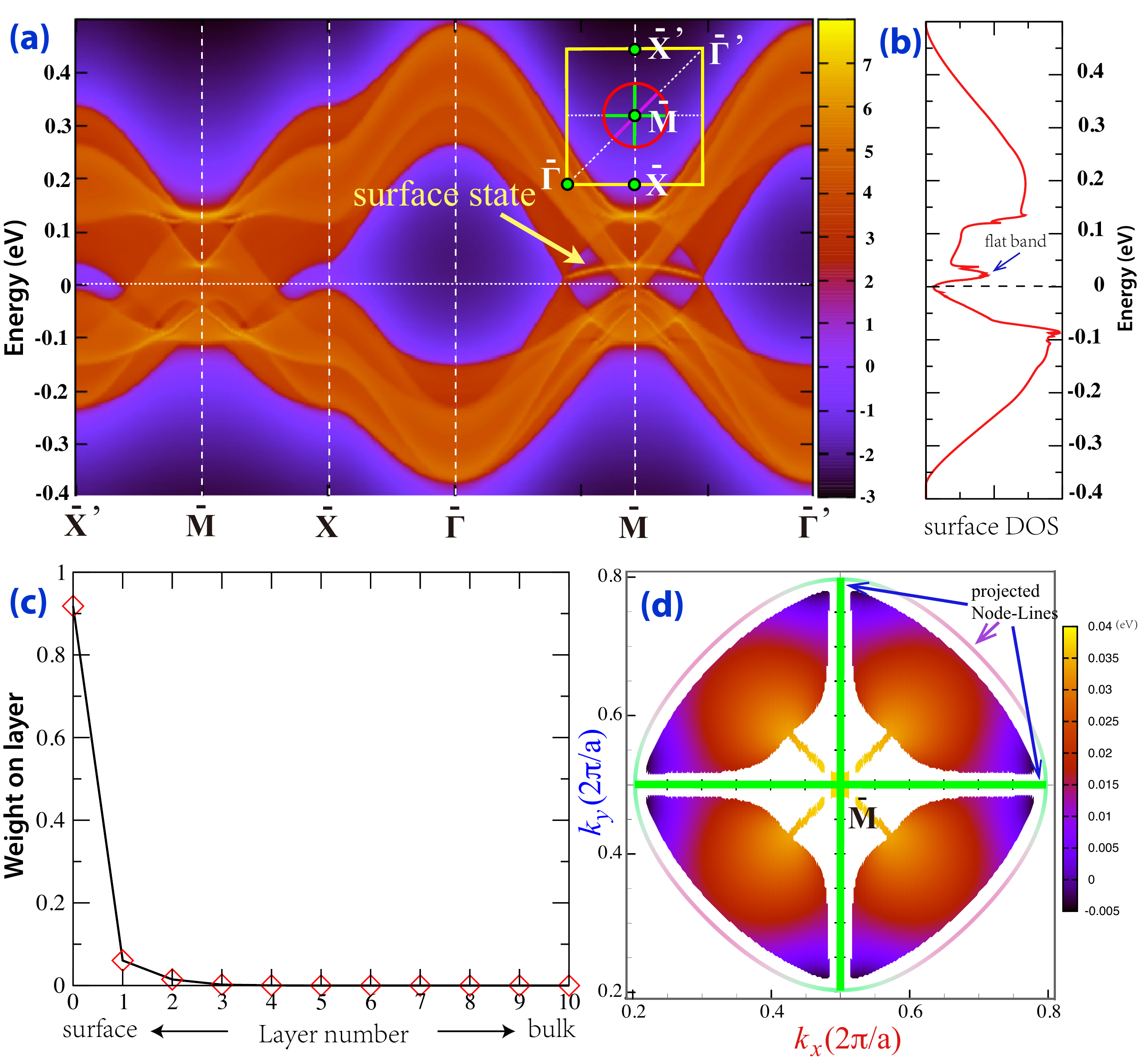}
\caption{(Color online) The (001)-surface state. (a) The nearly flat surface band is nestled between two solid Dirac cones, 
which are the projection of one of the node-line circles as indicated in the inset (red circle). The other two node-line rings 
are projected as two orthogonal diameters (green line). (b) The surface density of state. (c) The wave function of surface state pointed
by the arrow decay rapidly into bulk. (d) The eigen energy distribution of surface flat band nestled inside of projected node-line circle, 
which looks like a vibration model of ``drumhead". The mixing of surface and bulk state leads to discontinuous in this plot.
}
\label{surf}
\end{figure}

\subsection{Fermi surface and surface flat band}
 
 The two crossing bands within the $k_z$=$\frac{\pi}{a}$ plane obtained by TB Hamiltonian 
 are plotted in Fig. 3.
 In general the crossing of bands
do not happen at the same energy. They have energy dispersion around 25 meV. 
The alternative electron and hole pockets are formed when the band crossing is lower or higher 
than the Fermi level and this results in lotus-root like Fermi surface instead of 
dispersionless line. 

This topologically stable node-line semimetal state can have nontrivial 
surface states.~\cite{Ryu_2002PRL, JETP932011, JETP942011, 2011arXiv1111.4627V, burkov} For the (001)-surface, the three node-line rings are projected to 
be a ring and two orthogonal diameter segments inside of it as shown in Fig. 4(a). 
The (001)-surface state is calculated based on the six-band TB model using both Green's function method and slab-model method.~\cite{MRS_weng:9383312}
There is a nearly flat surface band nestled inside of the projected node-line ring with its band width 
being about 40 meV due to the particle-hole asymmetry. The peak-like surface density of states contributed by this nearly flat band is clearly shown in Fig. 4(b),
which is proposed to be an important route to high temperature surface superconductivity.~\cite{PhysRevB.83.220503, 2014arXiv1409.3944V} 
The layer-resolved weight of wave function for the surface flat band is shown in Fig. 4(c). It penetrates
just three layers into bulk with most of the weight on the surface layer. The surface localization of these flat bands 
is well resolved for those separated from bulk bands. The nestled flat surface states have small dispersion and 
their eigen energy distribution in surface BZ is shown in Fig. 4(d), which looks like some vibrational mode of ``drumhead". 
Such ``drumhead"-like states are readily to be detected by angle-resolved photoelectron spectroscopy or scanning tunnel microscope.

 The topological node-line state, as well as its surface flat band, can be
 understood by studying an effective 2$\times$2 toy model Hamiltonian. Taking $d_x$=$k_z$, 
 $d_y$=0 and $d_z$=$M-B(k_x^2+k_y^2+k_z^2)$, the Hamiltonian gives a node-line
 determined by $k_x^2+k_y^2=\frac{M}{B}$ in the plane $k_z$=$0$. 
 Obviously $\frac{M}{B} >0$ is required. The topology of this effective 
 continuum bulk hamiltonian has been analyzed~\cite{mong} 
 (See Appendix for details) and found to have topologically 
 protected (001) surface states with dispersionless zero eigen energy inside 
 of the projected node-line circle given by $\bar{k}_x^2+\bar{k}_y^2=\frac{M}{B}$.
 Here ($\bar{k}_x$, $\bar{k}_y$) denotes the k-point in (001) surface Brillouin zone. 
 As mentioned above, $d_0(\vec{k})$ determines the energy dispersion of the node-line,
 as well as the surface flat band, though the detailed dispersion of surface states are also 
 influenced by the surface potential in practice.~\cite{MRS_weng:9383312}

\section{Discussion}

We find that 6-1-1-p is not the only MTC having such novel node-line semimetal state. The 
MTC with the structure labeled as 6-1-2-p~\cite{ours} also has such nontrivial topological state.
(as shown Appendix) The differences are: (1) The band inversion happens at 
M point and the $Z_2$ index is (1;000) when even weaker SOC splitting (about 0.03 meV 
compared with 0.136 meV in 6-1-1-p) is considered. (2) The low energy physics around 
Fermi level can be described by six atomic like molecular orbitals also, but they are 
T$_{1u}$ ($p_x$, $p_y$ and $p_z$) and T$_{2g}$ ($d_{xy}$, $d_{yz}$ and $d_{xz}$). 
The similar tight-binding model on simple cubic lattice can also reproduce all of its electronic 
structure. (3) There are also three mutually perpendicular line-node circles centering M 
point instead of R point. The similar surface state with nearly flat band can also be obtained. 
Therefore, it is most plausible that there are more 3D MTCs which can host such node-line semimetal state.
Similar node-lines have also been found in optimally tuned photonic crystal composed of gyroid,~\cite{lufu} 
the Schwarz minimal G-surface. Other proposed carbon system include Bernal 
graphite~\cite{GP_Mikitik_2006PRB,GP_Mikitik_2008LTP} and 
hyper-honeycomb lattice.~\cite{2014arXiv1408.5522M} A carbon gyroid~\cite{PhysRevB_Gsurf} is 
found to be metal with Dirac cone in conduction bands higher away from Fermi level. Node-line is
also proposed in Dirac or Weyl superconductors.~\cite{FanZhang_2014PRL}

\section{Conclusion} \label{Conclusion}
 In summary, based on the first-principles calculations, we have
 predicted that a family of all-carbon 3D allotrope, Mackay-Terrones
 crystals, can have nontrivial topological node-line semimetal state,
 which is protected by both time-reversal symmetry and inversion
 symmetry after band inversion. When such bulk node-line
 is projected onto surface to form a circle, there is flat bands nestled inside of it. 
 Such ``drumhead"-like state is an ideal playground for many interaction induced 
 nontrivial states, such as superconductivity and fractional topological insulator states. 
 Further, if the inversion symmetry is broken, the node-lines will evolve into
 stable 3D Dirac points. Two examples of such MTC with stable structure have 
 been discussed. These predications can most probably be directly observed 
 in further experiments.








\section{acknowledgments}
H.M.W., Z.F. and X.D. acknowledge the supports from National Natural Science Foundation of China, the 973 program of
China (No. 2011CBA00108 and 2013CB921700) and the "Strategic Priority Research Program (B)" 
of the Chinese Academy of Sciences (No. XDB07020100). H.M.W. thanks the hospitality during his
stay in Tohoku University and part of this work has been done there. Y.K. acknowledges to 
the Russian Megagrant project grant No. 14.B25.31.0030. Both Y.L. and Y.K. are
supported by JST, CREST, ``A mathematical challenge to a new phase of material sciences" (2008-2013).

{\it Note:} During the reviewing of this work, we noticed that a similar work from Y. Chen {\it et al}.~\cite{2015arXiv150502284C}, 
in which the Node-line, the nestled nearly flat surface bands and the stable 3D Dirac nodes due to inversion symmetry breaking 
proposed in this manuscript are also obtained for another carbon system.

\bibliographystyle{unsrt}
\bibliography{mtc_ref}

\begin{appendix}
\section{Tight-binding Model} 
\subsection{6-1-1-p case}
The hypothetic atomic orbital basis set is arranged in the order of 
$g_{xy(x^2-y^2)}$, $g_{yz(y^2-z^2)}$, $g_{zx(z^2-x^2)}$, 
$f_{x(y^2-z^2)}$, $f_{y(z^2-x^2)}$ and f$_{z(x^2-y^2)}$. Those 
 from $g$ ($f$) orbitals are triple-degenerate and the on-site energy
 is set as $E_g$ ($E_f$). Since the cubic symmetry, only $g_{xy(x^2-y^2)}$ 
 orbital is plotted in the $xy$ plane in Figure S1. 
 The $f_{x(y^2-z^2)}$ is plotted as two parts $f_{xy^2}$ and 
 $-f_{xz^2}$, perpendicular to each other. 
 Arranging these orbitals on a simple cubic lattice with 
 lattice constant $a$, the Slater-Koster parameters for nearest-neighbor 
 hopping is defined in the following. The nearest hopping between $g_{xy(x^2-y^2)}$ 
 in $x$ and $y$ direction is $V_{ggp}$, while that in $z$ direction is $V_{ggd}$. 
 The hopping between $f_{xy^2}$ ($-f_{xz^2}$) along $x$ and $y$ ($x$ and $z$)
 is $V_{ffp}$, and that along $z$ ($y$) direction is $V_{ffd}$. The hopping 
 between nearest neighboring $g_{xy(x^2-y^2)}$ and $f_{xy^2}$ ($-f_{xz^2}$)
 along $y$ direction is $V_{fgp}$ ($V_{fgd}$), while those along 
 $x$ and $z$ direction are zero. We list some of the nonzero elements 
 of final tight-binding hamiltonian and others can be easily derived by 
 using the cubic cyclic symmetry.
 \begin{eqnarray*}
 H_{g_{xy(x^2-y^2)},g_{xy(x^2-y^2)}}=E_g+2 \cos(\vec{k}\cdot \vec{a}_x)V_{ggp} \\
 + 2 \cos(\vec{k}\cdot \vec{a}_y)V_{ggp}+2 \cos(\vec{k}\cdot \vec{a}_z)V_{ggd}
 \end{eqnarray*}
 \begin{eqnarray*}
 H_{g_{xy(x^2-y^2)},f_{x(y^2-z^2)}}=i*2 \sin(\vec{k}\cdot \vec{a}_y)V_{fgp} \\
 + i*2 \sin(\vec{k}\cdot \vec{a}_y)V_{fgd}
 \end{eqnarray*}
 \begin{eqnarray*}
 H_{g_{xy(x^2-y^2)},f_{y(z^2-x^2)}}=i*2 \sin(\vec{k}\cdot \vec{a}_x)V_{fgp} \\
 + i*2 \sin(\vec{k}\cdot \vec{a}_x)V_{fgd}
 \end{eqnarray*}
  \begin{eqnarray*}
 H_{f_{x(y^2-z^2)},f_{x(y^2-z^2)}}=E_f \\
+2 \cos(\vec{k}\cdot \vec{a}_x)V_{ffp}  + 2 \cos(\vec{k}\cdot \vec{a}_x)V_{ffp} \\
-2 \cos(\vec{k}\cdot \vec{a}_y)V_{ffp}  - 2 \cos(\vec{k}\cdot \vec{a}_y)V_{ffd} \\
-2 \cos(\vec{k}\cdot \vec{a}_z)V_{ffp}  - 2 \cos(\vec{k}\cdot \vec{a}_z)V_{ffd} 
 \end{eqnarray*}
Here $\vec{a}_x$, $\vec{a}_y$ and $\vec{a}_z$ are the nearest neighbor site along
positive $x$, $y$ and $z$ direction, respectively. We have fitted all the Slater-Koster 
parameters and find that $E_g$=-0.12, $E_f$=0.19, $V_{ffp}$=0.019, $V_{ffd}$=-0.075,
$V_{fgp}$=0.05, $V_{fgd}$=0.0, $V_{ggp}$=-0.035, $V_{ggd}$=-0.055 (all in unit eV)
can well reproduce the band structure from first-principles calculation as shown in Fig. 2 
of main text.

However, the following set of parameters will modify the band structure by shifting
the band crossing from R-M to $\Gamma$-M. The band structure with $E_g$=-0.10, $E_f$=0.16, 
$V_{ffp}$=-0.010, $V_{ffd}$=-0.080, $V_{fgp}$=0.05, $V_{fgd}$=0.0, $V_{ggp}$=-0.055, 
$V_{ggd}$=-0.035 is shown in Figure S2. Compared with that in realistic case, 
there is additional band inversion at M. This changes the Z$_2$ index to be (0;111) if the tiny 
SOC is considered.

\begin{figure}[tbp]
\includegraphics[clip,scale=0.35,angle=0]{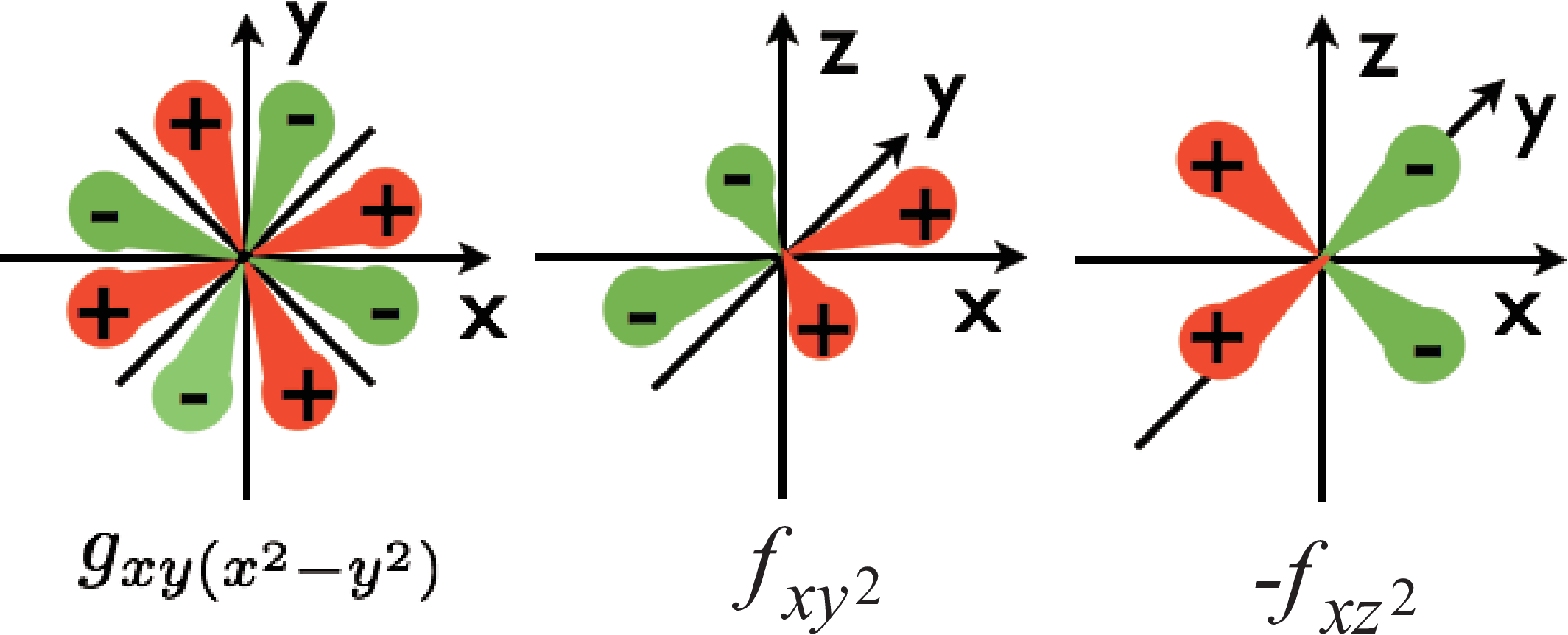}
\\
{\bf Figure S1}. The angular distribution of $g_{xy(x^2-y^2)}$, $f_{xy^2}$ and $-f_{xz^2}$ orbitals.
\label{g_f_orbital}
\end{figure}

\begin{figure}[tbp]
\includegraphics[clip,scale=0.35,angle=0]{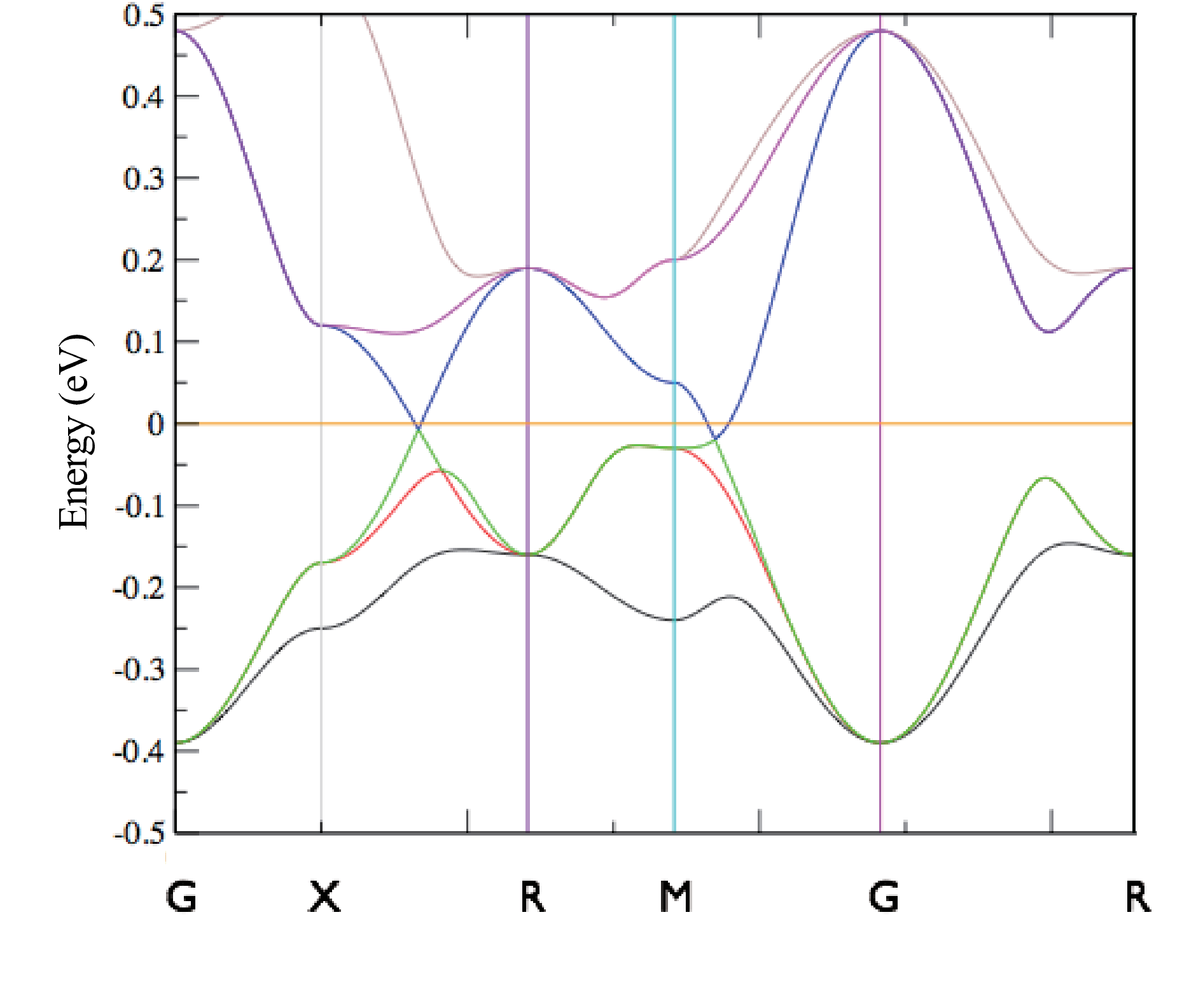}
\\
{\bf Figure S2}. The tight-binding band structure with $E_g$=-0.10, $E_f$=0.16, $V_{ffp}$=-0.010, $V_{ffd}$=-0.080,
$V_{fgp}$=0.05, $V_{fgd}$=0.0, $V_{ggp}$=-0.055, $V_{ggd}$=-0.035. The band crossing point is shifted from R-M to M-$\Gamma$ compared with Fig. 3 in main text.
\label{tbband2}
\end{figure}

\subsection{6-1-2-p case}
For 6-1-2-p case, the band structure from first-principles calculation is shown in Figure S3. 
Obviously, there are band inversion around M point. Careful analysis has shown that 
the occupied and unoccupied triply degenerated bands at $\Gamma$ (also at R point) 
are $T_{1u}$ and $T_{2g}$, respectively. Therefore, we can take hypothetic $p_x$, $p_y$ and $p_z$ 
orbitals as the basis for $T_{1u}$ representation in cubic symmetry and 
$d_{xy}$, $d_{yz}$ and $d_{zx}$ form $T_{2g}$. All the $p$ orbitals
have on-site energy $E_p$ and $d$ orbitals have $E_d$. The Slater-Koster parameters 
such as $V_{pp\sigma}$, $V_{pp\pi}$, $V_{dd\pi}$, $V_{dd\delta}$ and $V_{pd\pi}$ 
are defined as in common.~\cite{SKTB} Putting these orbitals on a simple cubic lattice, we can
have the tight-binding Hamiltonian and some of the nonzero elements are listed as following:
 \begin{eqnarray*}
 H_{x,x}=E_p+2\cos(\vec{k}\cdot \vec{a}_x)V_{pp\sigma} \\ 
 +2\cos(\vec{k}\cdot \vec{a}_y)V_{pp\pi}+2\cos(\vec{k}\cdot \vec{a}_z)V_{pp\pi} 
  \end{eqnarray*}
 \begin{eqnarray*}
 H_{x,xy}= i*2\sin(\vec{k}\cdot \vec{a}_y)V_{pd\pi}
 \end{eqnarray*}
 \begin{eqnarray*}
 H_{x,zx}= i*2\sin(\vec{k}\cdot \vec{a}_z)V_{pd\pi}
 \end{eqnarray*}
 \begin{eqnarray*}
 H_{xy,xy}= E_d +2\cos(\vec{k}\cdot \vec{a}_x)V_{dd\pi} \\
 +2\cos(\vec{k}\cdot \vec{a}_y)V_{dd\pi}+2\cos(\vec{k}\cdot \vec{a}_z)V_{dd\delta}
 \end{eqnarray*}
The other elements can be obtained by using cubic cyclic symmetry. The fitted parameters 
which can well reproduce the band structure from first-principles calculation are 
$E_p$=-0.10147, $E_d$=0.28281, $V_{pp\sigma}$=0.02005, $V_{pp\pi}$=-0.017848, 
$V_{pd\pi}$=0.034711, $V_{dd\pi}$=0.04694, $V_{dd\delta}$=-0.062523. The mean
square error, around 0.0061 eV$^2$, is estimated for the three conduction bands. Estimation for the
valence bands has problem in selecting proper bands since there are more than three bands
entangled. While the topology of bands from fitted model is the same as from first-principles calculation.

\begin{figure}[tbp]
\includegraphics[clip,scale=0.40,angle=0]{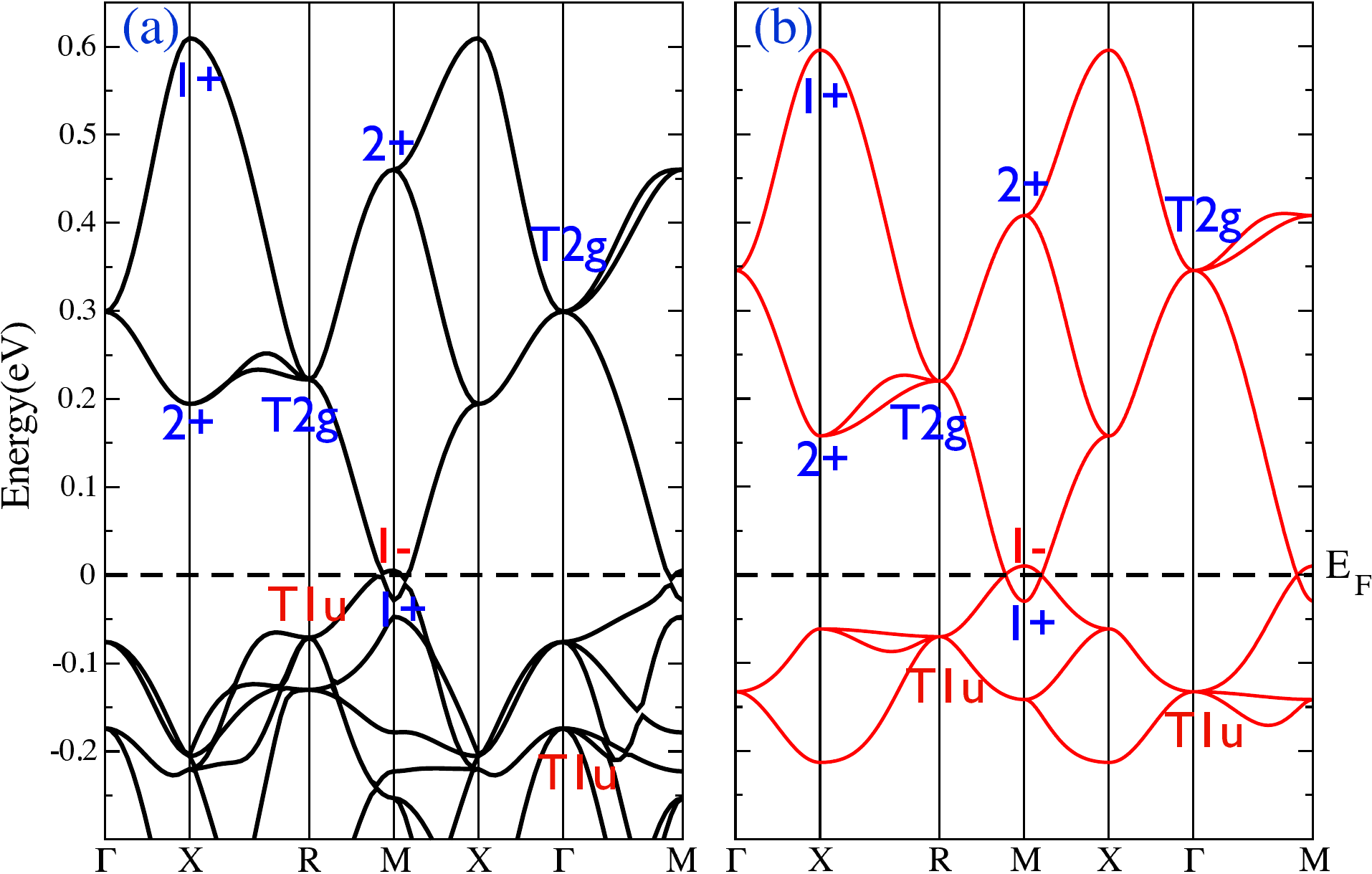}
\\
{\bf Figure S3}. The band structure of 6-1-2-p case calculated from (a) first-principles and (b) tight-binding model, respectively. The
symmetrical representation, degeneracy and parity of relevant bands are labeled.
\label{612pband}
\end{figure}

 \section{Real Valued Hamiltonian for Spinless System with Both Time-reversal and Inversion Symmetry}
 We will show that for a spineless system with both time-reversal (T) and inversion (P) symmetry its Bloch Hamiltonian $H(\mathbf{k})$
 can always be taken as real valued under some certain gauge choice. 
 To be general, plane waves $e^{i(\mathbf{k}+\mathbf{G}_n)\cdot \mathbf{r}}$ are taken as basis set
 to describe $H(\mathbf{k})$, where $\mathbf{G}_n$ are the reciprocal lattice vectors. 
 With the above chosen basis set, the invariance under time reversal operator $T$ can be expressed as,
 
  \begin{eqnarray*}
 \hat{T} H(\mathbf{k})\hat{T}^{-1} = \hat{O}H^*(\mathbf{k})\hat{O}^{-1}=H(\mathbf{-k})
 \end{eqnarray*} 
 and that of the inversion symmetry reads
   \begin{eqnarray*}
  \hat{O}H(\mathbf{k})\hat{O}^{-1}=H(\mathbf{-k})
 \end{eqnarray*} ,
 where the unitary matrix $\hat{O}$ can be defined as $\hat{O}_{nm}=1$ for $n=-m$ and zero for all the other matrix 
 elements. From the above two equations, it is obvious that $H(\mathbf{k})$ is real.

 \section{Topology of bulk Node-line Hamiltonian} 
As shown in main text, the effective hamiltonian for bulk Node-line state can be written as
 \begin{eqnarray*}
H(k_x, k_y, k_z)=\mathbf{d}(k_x, k_y, k_z)\cdot \mathbf{\sigma}
 \end{eqnarray*}
 , where $\mathbf{d}$=$(d_x, d_y, d_z)$ and $\mathbf{\sigma}=(\sigma_x,\sigma_y,\sigma_z)$.
 Taking $d_x$=$k_z$, $d_y$=0 and $d_z$=$M-B(k_x^2+k_y^2+k_z^2)$ can reproduce
 the bulk node-line state when $\frac{M}{B} >0$. According to R. S. K. Mong et al.~\cite{mong}, 
 to check its boundary state in surface perpendicular to $k_z$ this bulk hamiltonian should be reformulated as
 \begin{eqnarray*}
H(k_\|, k_z)=\mathbf{c}_0+\mathbf{c}_1 k_z+\mathbf{c}_2 k_z^2 \\
 =(0,0,M-Bk_\|^2)+(1,0,0)k_z+(0,0,-B)k_z^2.
 \end{eqnarray*}
Here $k_\|$ denotes in-plane coordinates $(k_x, k_y)$ and $\mathbf{c}_0$, $\mathbf{c}_1$ and $\mathbf{c}_2$ 
are vectors in space spanned by $\sigma$. The above bulk hamiltonian is parabola in plane 
spanned by $\mathbf{c}_1$ and $\mathbf{c}_2$. Its origin is within the concave side of the parabola
when $k_\| (k_x, k_y)$ taking the value satisfying $k_x^2+k_y^2<\frac{M}{B}$. Thus, there is 
topologically protected surface states nestled inside of projected node-line with 
zero eigen energy to form ``drumhead"-like state.
\end{appendix}

\end{document}